\documentclass[conference]{IEEEtran}
\IEEEoverridecommandlockouts

\usepackage{amsmath}
\usepackage{graphicx}
\usepackage{booktabs}
\usepackage{array, makecell}
\usepackage{multicol}
\usepackage{multirow}
\usepackage{amssymb}
\usepackage{pifont}
\usepackage[table,xcdraw]{xcolor}
\usepackage{algorithm,algpseudocode}
\usepackage{pgfplots}
\usepackage{pgfplotstable}
\usepackage{tikz}
\usepackage{cite}
\usepackage{hyperref}

\DeclareMathOperator*{\argmin}{argmin}

\newcommand*{\argminl}{\argmin\limits}

\definecolor{turquoise}{cmyk}{0.65,0,0.1,0.3}
\definecolor{purple}{rgb}{0.65,0,0.65}
\usepackage{textcomp}
\usepackage{xcolor}
\def\BibTeX{{\rm B\kern-.05em{\sc i\kern-.025em b}\kern-.08em
    T\kern-.1667em\lower.7ex\hbox{E}\kern-.125emX}}

\begin{document}

\newcommand{\shortname}{StableQuant}
\title{StableQuant: Layer Adaptive Post-Training Quantization for Speech Foundation Models}

\author{\IEEEauthorblockN{Yeona Hong, Hyewon Han, Woo-jin Chung, and Hong-Goo Kang}
\IEEEauthorblockA{
\textit{Department of Electrical and Electronic Engineering, Yonsei University} \\
Seoul, Republic of Korea \\
{\{yeonahong, hwhan, woojinchung\}}@dsp.yonsei.ac.kr, hgkang@yonsei.ac.kr}
}
\maketitle
\begin{abstract}
In this paper, we propose~\shortname, a novel adaptive post-training quantization (PTQ) algorithm for widely used speech foundation models~(SFMs). 
While PTQ has been successfully employed for compressing large language models~(LLMs) due to its ability to bypass additional fine-tuning, directly applying these techniques to SFMs may not yield optimal results, as SFMs utilize distinct network architecture for feature extraction.
\shortname~demonstrates optimal quantization performance regardless of the network architecture type, as it adaptively determines the quantization range for each layer by analyzing both the scale distributions and overall performance.
We evaluate our algorithm on two SFMs, HuBERT and wav2vec2.0, for an automatic speech recognition (ASR) task, and achieve superior performance compared to traditional PTQ methods.
\shortname~successfully reduces the sizes of SFM models to a quarter and doubles the inference speed while limiting the word error rate (WER) performance drop to less than $0.3\%$ with 8-bit quantization.
\end{abstract}


\begin{IEEEkeywords}
Post-training quantization, speech foundation models, model compression
\end{IEEEkeywords}

\section{Introduction}
\label{sec:intro}
Speech foundation models (SFMs) are neural networks trained to extract speech characteristics in a self-supervised manner using large-scale speech datasets~\cite{ssl_review}. 
They effectively capture high-level information embedded in speech, making them versatile for a wide range of downstream tasks~\cite{superb}, including recognition~\cite{schneider19_interspeech}, enhancement~\cite{ssl_enh}, and generation~\cite{speartts,soundstream}.
To capture the complex nature of speech, most SFMs are constructed with deep stacks of convolutional layers and often incorporate Transformers~\cite{vaswani2017attention} to model temporal sequences.
However, these architectures require a vast number of trainable parameters and significant computational resources, making them challenging to deploy on low-spec GPUs or in real-time applications. 
Consequently, compressing SFMs while preserving their performance remains a significant challenge.

To tackle this issue, numerous studies have explored the application of model compression techniques~\cite{gholami2022survey}, such as knowledge distillation~\cite{distillhubert2022,LightHuBERT2022} and weight pruning techniques~\cite{hjpruning2023} to foundation models.
However, these approaches necessitate the design and training of student networks or further fine-tuning, which is both time- and resource-intensive. 
Meanwhile, several studies in the natural language processing domain have successfully reduced the complexity of large language models (LLMs)~\cite{bert} using post-training quantization (PTQ) schemes~\cite{banner2019post, cai2020zeroq, nagel2021white}, which quantize pre-trained models without the need for additional fine-tuning.
These methods are designed to preserve the essential properties of LLMs, with some even achieving 1-bit quantization~\cite{ma2024era}.

Building on the success of quantization in LLMs, we aim to efficiently compress SFMs using a PTQ scheme, which has been relatively underexplored in the context of speech foundation models.
While PTQ has been successfully applied to LLMs~\cite{smoothquant2023}, directly transferring this approach to SFMs may lead to performance degradation due to the differing characteristics and architectures. 
LLMs are designed to process discrete token representations of words, which are inherently coarse, whereas speech signals are continuous and exhibit strong temporal correlations.
Therefore, additional pre-processing modules, such as convolution neural networks (CNN), are commonly employed before inputting data into sequential modules in the speech domain.
These differences often result in distinct distributions of weights or activations across layers, making it essential to design a tailored quantization strategy.
Notably, the distribution of the activations in CNN
tends to be significantly wider—up to 100 times compared to that of Transformers.
This broader range makes CNN more prone to outliers, which can adversely affect quantization.

In this paper, we propose \shortname, a layer-wise adaptive calibration method that optimally determines the quantization range for SFMs by preemptively removing redundant outliers, while considering each layer's unique characteristics. 
Specifically, we employ a percentile-based technique to selectively remove outliers on a layer-by-layer basis.
After removing the outliers, we perform quantization and dequantization, followed by calculating the mean squared error (MSE) to evaluate the resulting accuracy.
This adaptive approach to MSE calculation, which emphasizes outlier removal before quantization, offers an improvement over traditional methods that compute MSE over the entire range without considering outliers.
Through our method, we aim to enhance quantization accuracy and better maintain overall model performance.

To the best of our knowledge, this is the first work to apply PTQ to speech foundation models. 
Experimental results in automatic speech recognition (ASR) show that our proposed quantization scheme successfully compresses models while preserving performance, without the need for additional fine-tuning.
Moreover, we show the benefits of PTQ by demonstrating accelerated hardware performance with significantly fewer storage requirements, showing its potential for academic and industrial applications. 
\section{Related Works}
\subsection{Speech foundation models~(SFMs)}
characteristics via self-supervised learning approaches~\cite{chen2022wavlm}.
In this work, we consider two widely used SFMs: wav2vec2.0~\cite{baevski2020wav2vec} and HuBERT~\cite{hsu2021hubert}.
Wav2vec2.0 is trained to extract contextual information over longer temporal ranges to compensate for the corrupted segments of a given speech signal.
HuBERT leverages an offline clustering step to provide target labels for a masked language modeling (MLM)-like objective.
SFMs usually utilize a convolutional neural network (CNN)-based feature extractor along with Transformer layers to effectively model speech characteristics.

\subsection{Post-training quantization~(PTQ)}
Quantization~\cite{wu2020integer,krishnamoorthi2018quantizing} involves converting values to a lower-precision formats to reduce memory usage and enhance computational speed. 
Formally, a simple formulation for quantization function $F_q$ for a scalar input value $x$ can be defined as follows:
\begin{equation}
F_q(x,s,z) = \text{round}\left(\frac{x}{s}\right)+z,
\label{eq:quant}
\end{equation}
where $s$ and $z$ denote the scaling factor and zero-point offset, respectively.
The scaling factor controls quantization granularity, while zero-point offsets shift the quantized values to the desired range.
Both parameters are critical for quantization performance, affecting information retention and precision, making careful selection essential.
In this work, we use symmetric quantization ($z=0$), reducing the problem to determining the scale factor.


Since lower precision quantization can adversely impact model performance, the quantization algorithm must be designed to balance performance trade-offs.
To reduce quantization error, several calibration methods can be considered: Max, Percentile, Entropy, and MSE.
\textbf{Max} computes the scale factor by identifying the maximum absolute value from the distributions.
\textbf{Percentile} sets the range based on a percentile of absolute values to reduce the impact of extreme outliers. 
\textbf{Entropy} and \textbf{MSE} explore different quantization ranges by computing the error function (e.g., KL Divergence or mean-squared error) between the original floating-point distribution and candidate quantized distributions, enabling effective low-precision performance by selecting the minimum error. 
Each method has its strengths and weaknesses.
Max uses the full distribution range but is sensitive to outliers. 
Percentile improves resolution by removing outliers but risks information loss due to its relatively simplistic outlier selection.
MSE and Entropy are informative but depend on data distributions, that can be skewed by outliers. 
Thus, our approach aims to combine the strengths of the Percentile and MSE to mitigate these drawbacks.


\section{Proposed Methodology}
\vspace{-2pt}
\begin{figure}[t]
\centering
\begin{minipage}[t]{0.85\linewidth}
  \centering
  \centerline{\includegraphics[width=\columnwidth] {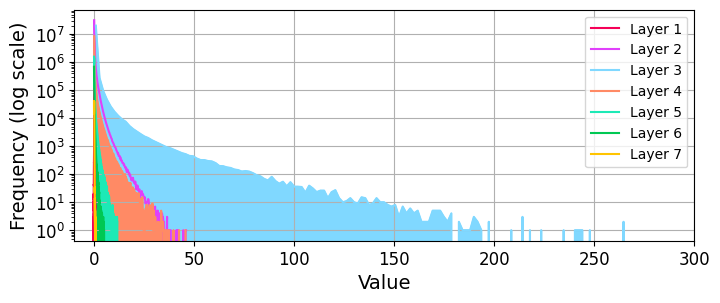}}
  \vspace{-5pt}
  \centerline{(a)}
    \label{fig:distribution_conv_layer}
    \vspace{5pt}
\end{minipage}
\hfill
\centering
\centering
\begin{minipage}[t]{0.85\linewidth}
  \centering
  \centerline{\includegraphics[width=\columnwidth] {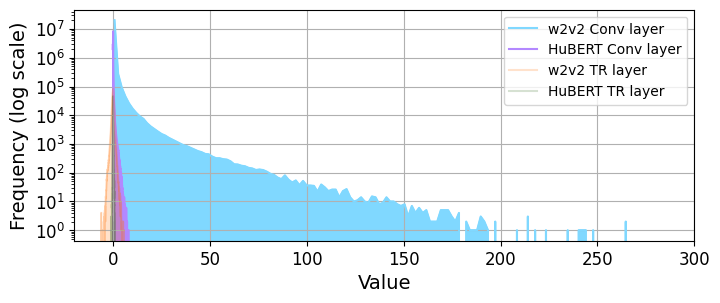}}
  \vspace{-5pt}
  \centerline{(b)}
  \label{fig:distribution_w2v2_hubert_act}
  \vspace{5pt}
\end{minipage}
\hfill
\centering
\begin{minipage}[t]{0.85\linewidth}
  \centering
  \centerline{\includegraphics[width=\columnwidth] {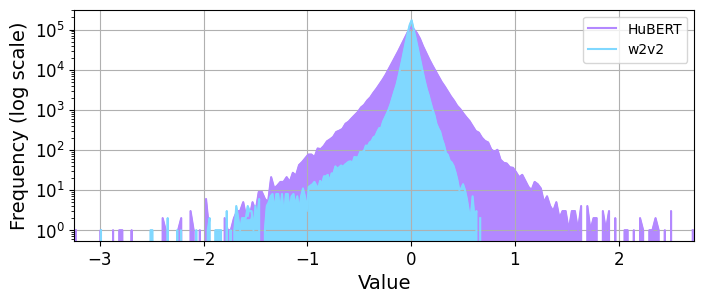}}
  \vspace{-5pt}
  \centerline{(c)}
  \label{fig:distribution_w2v2_hubert_weight}
\end{minipage}
\caption{(a) Amplitude distribution of activations for each convolution layer in wav2vec2.0. (b) Amplitude distributions of activations in wav2vec2.0 and HuBERT. (c) Amplitude distributions of weights in wav2vec2.0 and HuBERT.}

\label{fig:distribution}
\end{figure}

\setlength{\textfloatsep}{5pt}
\begin{algorithm}[!t]

\caption{\shortname~procedure for activations}
\label{algorithm}
\begin{algorithmic}[1]

\small\Statex \textbf{Input:} Number of layers $L$ with activations $\mathbb{A}=\left\{a^l\right\}_{l=1}^L$, speech foundation model $\mathbb{N}(\mathbb{A})=\left\{N(a^l)\right\}_{l=1}^L$, calibration/development dataset $D_{stat}$/$D_{dev}$, threshold $\gamma$, Cut-off percentile candidates $\mathbb{P}=\left\{p_k\right\}_{k=1}^K$, Scale calibration function $S_c(p,m;D)$, layer-wise quantization function $F_q(a^l, S_c)$ for $a^l$ using $S_c(p,m;D)$,  ASR evaluator $G(\mathbb{N}(\mathbb{A});D)$ 
\small\Statex \textbf{Output:} Set of layers to clip $\mathbb{S}$, Quantized neural network $\mathbb{N}_q(\mathbb{A})=\left\{N_q(a^l)\right\}_{l=1}^L$
\small\Statex \textbf{Stage 1:}. Layer Selection
\small\State Initialize $\mathbb{S}=\emptyset$ \Comment{Initialize clipping layer set}

\small\For{$l$ in $1, ..., L$}
\small\State Initialize $\mathbb{N}_q=\mathbb{N}$ \Comment{Initialize the network}
    \small\State $N_q(a^l) \gets F_q(a^l, S_c(p=0, m=0; D_{stat}))$

    \small\State $\Delta=G(\mathbb{N}_q(\mathbb{A});{D_{dev}})-G(\mathbb{N}(\mathbb{A});{D_{dev}})$ 
    \small\If{$\Delta > \gamma$}
        \small\State $\mathbb{S}=\mathbb{S} \cup \left\{l\right\}$ \Comment{Select layer to clip}
    \EndIf
\EndFor
\\
\small\Statex \textbf{Stage 2:}. Layer-adaptive Quantization 
\small\State Initialize $\mathbb{Y}=\emptyset$ \Comment{Initialize score set}
\small\State Initialize $\mathbb{N}_q=\mathbb{N}$ \Comment{Initialize the network}
\small\For{$p_k$ in $\mathbb{P}$}
\small\For{$l$ in $1, ..., L$}
    \small\If{$l \in \mathbb{S}$}
    \small\State $N_q(a^l) \gets F_q(a^l,S_c(p=p_k, m=1; D_{stat}))$
    \Else
    \small\State $N_q(a^l) \gets F_q(a^l,S_c(p=0, m=1; D_{stat}))$
    \EndIf
    \EndFor
    \small\State $y = G(\mathbb{N}_q(\mathbb{A});{D_{dev}})$ 
    \small\State $\mathbb{Y}=\mathbb{Y} \cup \left\{y\right\}$ 
\EndFor
\small\State $p_{opt} \gets \argminl_{\mathbb{P}}\mathbb{Y}$

\small\State $\mathbb{N}_q(a^\mathbb{S}) \gets F_q(a^\mathbb{S},S_c(p=p_{opt}, m=1; D_{stat}))$
\small\State $\mathbb{N}_q(a^{\mathbb{S}^c}) \gets F_q(a^{\mathbb{S}^c},S_c(p=0, m=1; D_{stat}))$


\small\State \Return $\mathbb{N}_q(\mathbb{A})$ \Comment{Optimized quantization model}
\end{algorithmic}

\end{algorithm}

\subsection{Distributions of weights and activations} 
\label{sec:distribution}
\vspace{-2pt}


There are several key differences between the quantization processes for LLMs~\cite{bert} and SFMs~\cite{baevski2020wav2vec,hsu2021hubert}.
Due to the high correlation between neighboring frames in speech data, SFMs employ CNN-based feature extractors, unlike LLMs.
The strong inductive bias of CNN-based feature extractors can result in activations that exhibit unnormalized distributions.
This can result in significant performance degradation when directly applying traditional quantization methods to SFMs.

Fig.~\ref{fig:distribution}(a) shows the distributions of activations for each CNN layer in wav2vec2.0, illustrating that there are significant variations in the dynamic range of each layer's activations.
Transformer activations show significantly smaller variance for both HuBERT and wav2vec2.0 (Fig.~\ref{fig:distribution}(b)).
These results can be attributed to differences in the normalization schemes and the architectural characteristics between CNN and Transformer, as well as between SFMs and LLMs.
Meanwhile, the models' weight distributions generally exhibit a normal distribution with some outliers (Fig.~\ref{fig:distribution}(c)), making them relatively easier to quantize.
Through these observations, we can surmise that quantifying the feature representations from certain convolutional layers in these models may be challenging, requiring a careful calibration scheme depending on the layer characteristics.
This, in turn, implies that an adaptive quantization method for each layer may be necessary to effectively quantize these models.


\begin{algorithm}[!t]

\caption{Scale calibration function $S_c(p,m;D)$}
\label{calib_algorithm}
\begin{algorithmic}[1]
\Statex \textbf{Input:} Set of values $\mathbf{X}$ from calibration data $D$, flag for MSE computation $m$, Cut-off percentile $p$, number of bits for quantization $b$.
\Statex \textbf{Output:} Computed scale factor $s$
\State $H_D(X)=\text{Histogram}(X)$
\State $H_{D,clip}(X)=\text{Clipping} \left( H_D(X), p \right)$ \Comment{Percentile clipping}
\State Let $\mathbb{C}=\left\{c_i\right\}_{i=1}^B$ as the center of each bins of $H_{D,clip}(X)$
\If{$m=1$} \Comment{MSE computation}
\For{$i$ in $1, ..., B$} 
\State $\hat{s}_i=\frac{c_i}{2^{b-1}-1}$
\For{$j$ in $1,....,B$} 
    \State $\hat{c}_{j, s_i} = \text{round}\left( \frac{c_j}{s_i} \right) \cdot {s_i}$.
    \EndFor
    \State $e_i=\frac{1}{B} \sum_{j=1}^{B} \left( c_{j} - \hat{c}_{j, s_i} \right)^2 \cdot H_{D,clip}(X;j)$ 

\EndFor
\State $opt = \argminl_{k}e_i$
\State $s = s_{opt}$
\Else
\State $s=\frac{c_B}{2^{b-1}-1}$  \Comment{Percentile computation}
\EndIf
\State \Return $s$ \Comment{Scale factor}


\end{algorithmic}

\end{algorithm}



\vspace{-5pt}
\subsection{\shortname}
To address the aforementioned issue, we propose \shortname, a layer-wise adaptive quantization scaling algorithm that provides stable performance during the quantization of SFMs.
The detailed process is outlined in Algorithm~\ref{algorithm}.
Our method first identifies the layers where significant performance degradation occurs when applying traditional quantization methods. 
This is determined by measuring the difference between the original model's performance and that of the model after applying naive percentile quantization on a given layer.
For layers where this difference exceeds a specified threshold, $\gamma$ (set as 0.25$\%$ WER drop in our experiments), we compute scale factors based on the layer characteristics, as detailed in Algorithm~\ref{calib_algorithm}.
For the selected layers, percentile-clipping is applied to the output activations before calculating the MSE quantization error. This involves searching for an optimal cut-off percentile $p_{opt}$ within the range $\mathbb{P}\in[0,0.5]$ at intervals of 0.01.
By clipping the histogram $H(X)$, outliers are removed during scale factor determination to reflect the dominant distribution.
This pre-processing step is crucial for stable quantization, as it accounts for both outlier removal and the data distribution.
For other layers that are not in $\mathbb{S}$, only MSE computation is used for searching scale factor to preserve the overall distribution. 
By adjusting the process for each layer, our method achieves stable quantization.

\begin{table}[!t]
\vspace{-5pt}
\centering

\caption{WERs~(\%) for selectively quantizing network layers (TR-Layer: Transformer layers only, All-Layer: all layers). Applied the Percentile calibration. W8A8 indicates 8-bit quantization for both weight (W) and activations (A).} 
\label{tab:naive_quant}
\resizebox{0.87\columnwidth}{!}{
\begin{tabular}{l||r|r|r|r}
\toprule
\multirow{2}{*}{\textbf{Precision}} & \multicolumn{2}{c|}{\textbf{HuBERT}} & \multicolumn{2}{c}{\textbf{wav2vec2.0}} \\
 & TR-Layer & All-Layer & TR-Layer & All-Layer\\ \midrule
Baseline & 2.16 & 2.16 & 2.78 & 2.78\\
W16A16 & 2.16 & 2.17 & 2.78 & 2.82 \\
W8A8 & 2.21 & \textbf{3.03} & 2.78 & \textbf{95.28} \\
W6A6 & 2.55 & \textbf{96.6} & 2.98 & \textbf{99.96} \\
W4A4 & 98.08 & 99.98 & 61.03 & $>$100 \\ \bottomrule

\end{tabular}
}
\end{table}

\begin{table*}[!t]
\centering
\caption{WERs~(\%) for various quantization methods when applying the proposed layer-adaptive calibration strategies.}
\label{tab:proposed}
\resizebox{0.8\linewidth}{!}{%
\begin{tabular}{l||l|l|c|rrrr}
\toprule
\multirow{2}{*}{\textbf{Model}}      & \multicolumn{1}{c|}{\multirow{2}{*}{\textbf{Precision}}} & \multicolumn{1}{c|}{\multirow{2}{*}{\textbf{Clipped Layers} ($\mathbb{S}$)}} & \multicolumn{1}{c|}{\multirow{2}{*}{\begin{tabular}[c]{@{}c@{}}\textbf{Cut-off}\\ \textbf{Ratio} ($p_{opt}$)\end{tabular}}} & \multicolumn{4}{c}{\textbf{WER} (\%)}                                                                                             \\
                            & \multicolumn{1}{c|}{}                              & \multicolumn{1}{c|}{}                                & \multicolumn{1}{c|}{}                                                                           & \multicolumn{1}{c}{\textbf{Percentile}} & \multicolumn{1}{c}{\textbf{MSE}} & \multicolumn{1}{c}{\textbf{Entropy}} & \multicolumn{1}{c}{\textbf{StableQuant}} \\ \midrule
\multirow{4}{*}{HuBERT}     & W16A16                                             & -                                                    & 0.0                                                                                             & 2.16                           & 2.16                    & 2.17                        & 2.16                            \\
                            & W8A8                                               & conv 0, 1, 2                                         & 0.2                                                                                             & 2.52                           & 2.56                    & 2.65                        & \textbf{2.35}                            \\
                            & W6A6                                               & conv 0, 1, 2                                         & 0.16                                                                                            & 55.55                          & 6.33                    & 10.84                       & \textbf{3.98}                            \\
                            & W4A4                                               & conv 0, 1, 2                                         & 0.0                                                                                             & \textgreater{}100              & 100                     & 93.41                       & 89.83                           \\ \midrule
\multirow{4}{*}{wav2vec2.0} & W16A16                                             & -                                      & 0.0                                                                                             & 2.78                           & 2.78                    & 2.82                        & 2.78                            \\
                            & W8A8                                               & conv 0, 1, 2, 3                                      & 0.11                                                                                            & 3.45                           & 5.18                    & 5.21                        & \textbf{3.07}                            \\
                            & W6A6                                               & conv 0, 1, 2, 3                                      & 0.21                                                                                            & 99.95                          & 60.06                   & 42.92                       & \textbf{6.10}                            \\
                            & W4A4                                               & conv 0, 1, 2, 3                                      & 0.05                                                                                            & \textgreater{}100              & 99.43                   & 99.95                       & 99.27                           \\ \bottomrule
\end{tabular}
}
\vspace{-5pt}
\end{table*}

\section{Experiments and Results}
We utilize the wav2vec2.0 and HuBERT-Large models from the \texttt{fairseq} repository\footnote{https://github.com/facebookresearch/fairseq}, pre-trained on 60k hours of the LibriLight dataset~\cite{librilight} and fine-tuned for the ASR task using the LibriSpeech-960h dataset~\cite{panayotov2015librispeech}.
To create a compact network from the quantized model using the proposed calibration method, we utilized NVIDIA's \texttt{TensorRT}\footnote{https://github.com/NVIDIA/TensorRT} engine libraries~\cite{migacz20178}.
We used a single NVIDIA GeForce RTX 3090 GPU for inference.
During the quantization process, we performed calibration and validation using the \texttt{dev-clean} subset of LibriSpeech, followed by final ASR evaluation on the \texttt{test-clean} subset.


 

\begin{table}[t]
\centering
\caption{Comparison with existing PTQ and QAT method in terms of Model size~(MB) and WERs~(\%) for various HuBERT models.}
\vspace{-5pt}
\label{tab:comparison}
\resizebox{0.85\columnwidth}{!}{%

\begin{tabular}{ll|l|r|r}
\toprule
\multicolumn{2}{c|}{\multirow{2}{*}{Model type}}                              & \multicolumn{1}{c|}{\multirow{2}{*}{Precision}} & \multicolumn{1}{c|}{\multirow{2}{*}{\begin{tabular}[c]{@{}c@{}}Model\\ size (MB)\end{tabular}}} & \multicolumn{1}{c}{\multirow{2}{*}{\begin{tabular}[c]{@{}c@{}}WER\\ (\%)\end{tabular}}} \\
\multicolumn{2}{c|}{}                                                         & \multicolumn{1}{c|}{}                           & \multicolumn{1}{c|}{}                                                                           & \multicolumn{1}{c}{}                                                                    \\ \midrule
\multicolumn{1}{l|}{Base}                   & \multirow{2}{*}{Full precision} & \multirow{2}{*}{W32A32}                         & 491                                                                                             & 6.33                                                                                    \\
\multicolumn{1}{l|}{Large}                  &                                 &                                                 & 1205                                                                                            & 2.16                                                                                    \\ \midrule
\multicolumn{1}{l|}{\multirow{3}{*}{Large}}  & AdaRound~\cite{nagel2020up}     &   \multirow{3}{*}{W8A8}   & -       & 2.66    \\
\multicolumn{1}{l|}{}                        & StableQuant                     &                            & 353     & 2.35  \\
\multicolumn{1}{l|}{}                       & QAT                             &                            & 357     & 2.34  \\ 
\bottomrule


\end{tabular}
}
\end{table}



%
\subsection{Quantization difficulties for SFMs}

To examine the challenges of quantization for SFMs, we compared the results of quantizing only the Transformer layers with those of quantizing all layers, including the CNN feature extractor.
Table~\ref{tab:naive_quant} presents the quantization results at various bit precisions for both the Transformer-only and all-layer configurations for HuBERT and wav2vec2.0.
Quantization with precision below INT8 (W8A8) significantly degrades performance in the ALL-Layer configuration.
This suggests that conventional quantization methods are inadequate for SFMs incorporating CNN layers, as elaborated in Section~\ref{sec:distribution}.
These findings support our hypothesis that employing a layer-adaptive calibration strategy is essential for maintaining robust performance after quantization.

\subsection{Efficiency of~\shortname}
Table~\ref{tab:proposed} shows the quantization performance of both \shortname~and conventional quantization methods.
The clipped CNN layers and cut-off ratios are selected by our~\shortname.
\shortname~outperforms conventional methods across all precision levels, particularly with precision below INT8 (W8A8). 
Although the quantization with W4A4 still shows high-performance degradation due to the inherent complexity of speech signals, \shortname~successfully maintains the performance in W6A6.
This demonstrates that robust quantization results can be achieved through accounting for both crucial factors, layer-dependent outlier removal, and scale factor selection based on carefully established criteria.

We further compare the performance of the proposed StableQuant with both PTQ and QAT approaches, and AdaRound~\cite{nagel2020up}, an existing post-training quantization method utilizing layer-wise optimization of weight rounding, as demonstrated in Table~\ref{tab:comparison}.
Our StableQuant employing PTQ surpasses the quantized results with AdaRound and achieves competitive performance to QAT, requiring a sufficient training dataset for fine-tuning.
This demonstrates the efficacy of our algorithm compared to the existing PTQ method and the advantages of utilizing low-resource datasets.

\subsection{Model size efficiency and inference speed improvement}
We measured the reduction in model size and the increase in inference speed achieved with our~\shortname.
Notably, as shown in Table~\ref{tab:comparison}, the quantized HuBERT-Large model with W8A8 precision requires less storage in terms of model size than the HuBERT-Base model, while still demonstrating superior performance.
This outcome highlights our ability to reduce model size without compromising quantization stability. 
Additionally, quantizing both weights and activations resulted in a substantial improvement in inference speed, achieving nearly a twofold increase, as shown in Fig.~\ref{fig:speed}.

\pgfplotsset{compat=1.17}

\begin{figure}[t]
\label{fig:speed}
\centering
\begin{tikzpicture}[scale=0.9]  
\begin{axis}[
    width=1.\columnwidth,  
    height=0.65\columnwidth,  
    xlabel={Audio length (s)},
    ylabel={Inference time (ms)~($\downarrow$)},
    legend pos=north west,  
    legend style={draw=none, fill=none, font=\scriptsize, align=right},  
    grid=major,
    xtick=data,
    ymin=0,
    ymax=450,
    mark size=2.5pt,
    label style={font=\footnotesize},  
    tick label style={font=\footnotesize},  
    ]

    \addplot[
        color=turquoise,
        mark=square,
        thick,
        ] coordinates {
        (1,4.47) (5,6.29) (10,11.44) (30,62.23) (60,213)
        };
    \addlegendentry{wav2vec2.0 (W8A8)}
    \addplot[
        color=turquoise,
        mark=triangle,
        thick,
        ] coordinates {
        (1,9.15) (5,14.32) (10,24.72) (30,97.02) (60,287.96)
        };
    \addlegendentry{wav2vec2.0 (W8A32)}    
    
    \addplot[
        color=turquoise,
        mark=diamond,
        thick,
        ] coordinates {
        (1,11.95) (5,20.47) (10,38.97) (30,129.32) (60,382.95)
        };
    \addlegendentry{wav2vec2.0 (W32A32)}
    \addplot[
        color=purple,
        mark=o,
        thick,
        ] coordinates {
        (1,4.88) (5,6.95) (10,12.79) (30,70.34) (60,241.44)
        };
    \addlegendentry{HuBERT (W8A8)}
    \addplot[
        color=purple,
        mark=+,
        thick
        ] coordinates {
        (1,8.91) (5,13.80) (10,24.03) (30,94.79) (60,287.23)
        };
    \addlegendentry{HuBERT (W8A32)}
    \addplot[
        color=purple,
        mark=x,
        thick,
        ] coordinates {
        (1,12.11) (5,20.76) (10,39.06) (30,143.98) (60,386.82)
        };
    \addlegendentry{HuBERT (W32A32)}
\end{axis}
\end{tikzpicture}
\vspace{-5pt}
\caption{Inference speed of SFMs (ms) vs. Audio length (s)}
\label{fig:speed}
\end{figure}
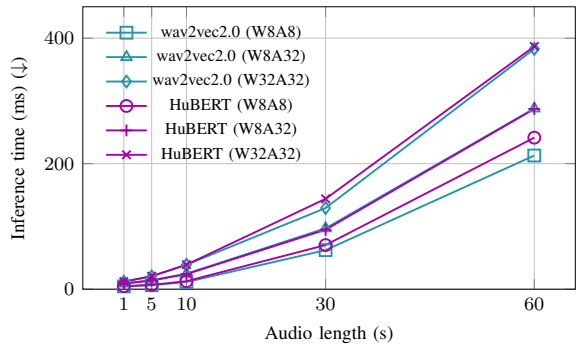

\section{Conclusion}
In this work, we proposed an efficient post-training quantization method for speech foundation models. To mitigate the impact of quantization on outlier parameters, we introduced an adaptive layer-wise scaling factor search algorithm that preserves model performance by identifying optimal scaling factors for core parameters while effectively determining the ranges for excluding outliers.
Extensive experimental results and analyses demonstrated the successful deployment of the proposed quantization scheme in automatic speech recognition applications, achieving performance comparable to baseline models even with quantization approaching 6 bits.
Our implementation is available in the repository for reproduction\footnote{https://github.com/Yeona-Hong/StableQuant}.

\vfill\pagebreak

\clearpage
 \bibliographystyle{IEEEbib}
 \bibliography{strings,refs}

\end{document}